\documentclass[]{elsart}
\usepackage{graphics}
\usepackage{graphicx}
\usepackage{epsfig}
\usepackage{amssymb}
\begin{document}
\begin{frontmatter}
\title{Analysis of high-resolution foreign exchange data of USD-JPY for 13 years}


\author{\small{Takayuki Mizuno\thanksref{fff}$^{a}$}},
\author{\small{Shoko Kurihara$^{a}$}},
\author{\small{Misako Takayasu$^{b}$}},
\author{\small{Hideki Takayasu$^{c}$}}

\address{$^{a}$Department of Physics, Faculty of Science and Engineering, Chuo University, Kasuga, Bunkyo-ku, Tokyo 112-8551, Japan}
\address{$^{b}$Department of Complex Systems, Future University-Hakodate, 116-2 Kameda-Nakano-cho, Hakodate, Hokkaido 041-8655, Japan}
\address{$^{c}$Sony Computer Science Laboratories Inc., 3-14-13 Higashigotanda, Shinagawa-ku, Tokyo 141-0022, Japan}

\thanks[fff]{Corresponding author.\\
{\it E-mail address:}\/ mizuno@phys.chuo-u.ac.jp (T.Mizuno)}

\begin{abstract}
We analyze high-resolution foreign exchange data consisting of 20 million 
data points of USD-JPY for 13 years to report firm statistical laws in 
distributions and correlations of exchange rate fluctuations. A conditional 
probability density analysis clearly shows the existence of trend-following 
movements at time scale of 8-ticks, about 1 minute.
\end{abstract}


\end{frontmatter}

% main text
\section{Introduction}
In the case of transactions of stocks, the places and times for trading are 
limited, but in the case of transactions of foreign exchanges dealers in the 
world are trading 24 hours except weekends. About 200 trillion yen is traded 
by the exchange market in the world in one day. This amount of money is 
equivalent to about 2.5 times the national budget of Japan for one year. Due 
to these continuous and gigantic properties the characteristics of the open 
market appear most notably in the foreign exchange.

The fat-tail distributions of price fluctuations [1], the diffusions of 
price [2,3], the correlations of price fluctuations [4,5,6] are well-known 
characteristics of the open market. However, some of them are not conclusive 
enough because the terms of data are not long enough. We analyze all the Bid 
record (about 20 million ticks) of the exchange rates for "Yen/Dollar" that 
were traded by the term from 1989 to 2002 to report the firm statistical 
laws. We use so-called CQG database in the present study. Fluctuations as 
large as 1.5 yen/dollar or more are observed in the high-frequency (= tick) 
data of CQG, but we have deleted them by the filter in the analysis of this 
paper because it is reasonably thought that these fluctuations are caused by 
manual errors.

\section{The statistical laws of Exchange rate fluctuations}
It is important to observe the distribution of price difference in a study 
of the dynamics of price fluctuations. In Fig. 1 we show the cumulative 
probability distributions of the absolute value of (a) positive ($dP(t)>0$ 
means weaker-yen {\&} stronger-dollar, because we observe the Yen/Dollar 
rate.) and (b) negative ($=$stronger-yen {\&} weaker-dollar) price 
fluctuations. Time scale unit (dt) is one minute, and each rate change 
distribution is normalized by the standard deviation. First, we focus the 
distribution in the time scale of 1 minute ($dt=1$). The dashed lines in Fig.1 
(a) and (b) are the normal distributions. It is found that the price change 
distribution for 1 minute has tails much fatter than those of the normal 
distribution. We can use the power distribution to approximate the fat tails 
of this distribution. In both cases of positive and negative differences, 
the power exponent of distribution for 1 minute time scale is about $-2.5$. 
These figures are quite symmetrical. Next we pay attention to the larger 
time scales. In the positive case the rate change distribution tends to 
converge to the normal distribution. When time scale (dt) is 10000 minutes 
(about 1 week), the rate change distribution is nicely approximated by the 
normal distribution. However, in the negative case the distributions keep 
the power law even in the case of ($dt=10000$) nearly 1 week in contrast to 
the positive case. Namely, the assumption of normal distribution which is 
commonly applied in the financial technology is not valid for the negative 
changes even in the case of large time scales. Therefore, many theories in 
financial technology have the inconsistency for all time scales. It should 
be noticed from these figures that the distribution of large time scales is 
more asymmetrical than the distribution of the short time scales. To clarify 
the asymmetry, we introduce the skewness, which is defined by third moments 
as described by the following equation,

\begin{equation}
\label{eq1}
\mbox{Skewness = }\frac{\left\langle {\left( {dP(t,dt) - \left\langle 
{dP(t,dt)} \right\rangle } \right)^3} \right\rangle }{\sigma ^3}
\end{equation}

where, $\sigma $ is the standard deviation and $dP(t,dt)$ is the price 
difference over time scale $dt$. The value of skewness is zero for symmetric 
case. In Fig.2 we show the scale dependence of the skewness. The skewness of 
the rate change distribution of short time scale is almost zero. However, 
the skewness shows the negative value for large time scale because in the 
case of large time scales the change of very strong yen often occurs. The 
largest asymmetry of the rate change distribution is observed with the time 
scale of $dt=3$ days.

\begin{center}
\begin{figure}
\begin{minipage}{.5\linewidth}
\centerline{\includegraphics[width=2.52in,height=1.54in]{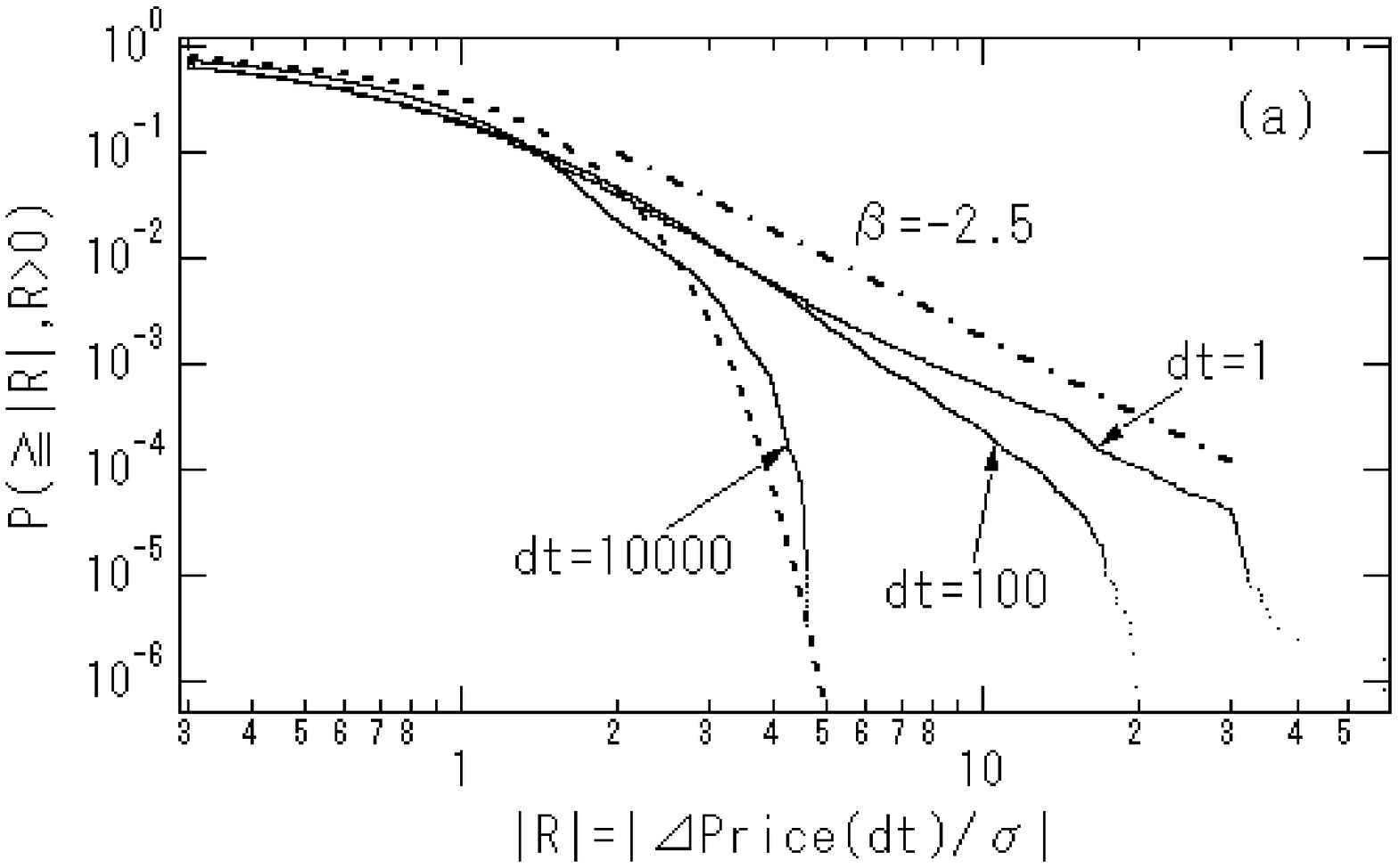}}
\label{fig1}
\end{minipage}
\begin{minipage}{.5\linewidth}
\centerline{\includegraphics[width=2.52in,height=1.54in]{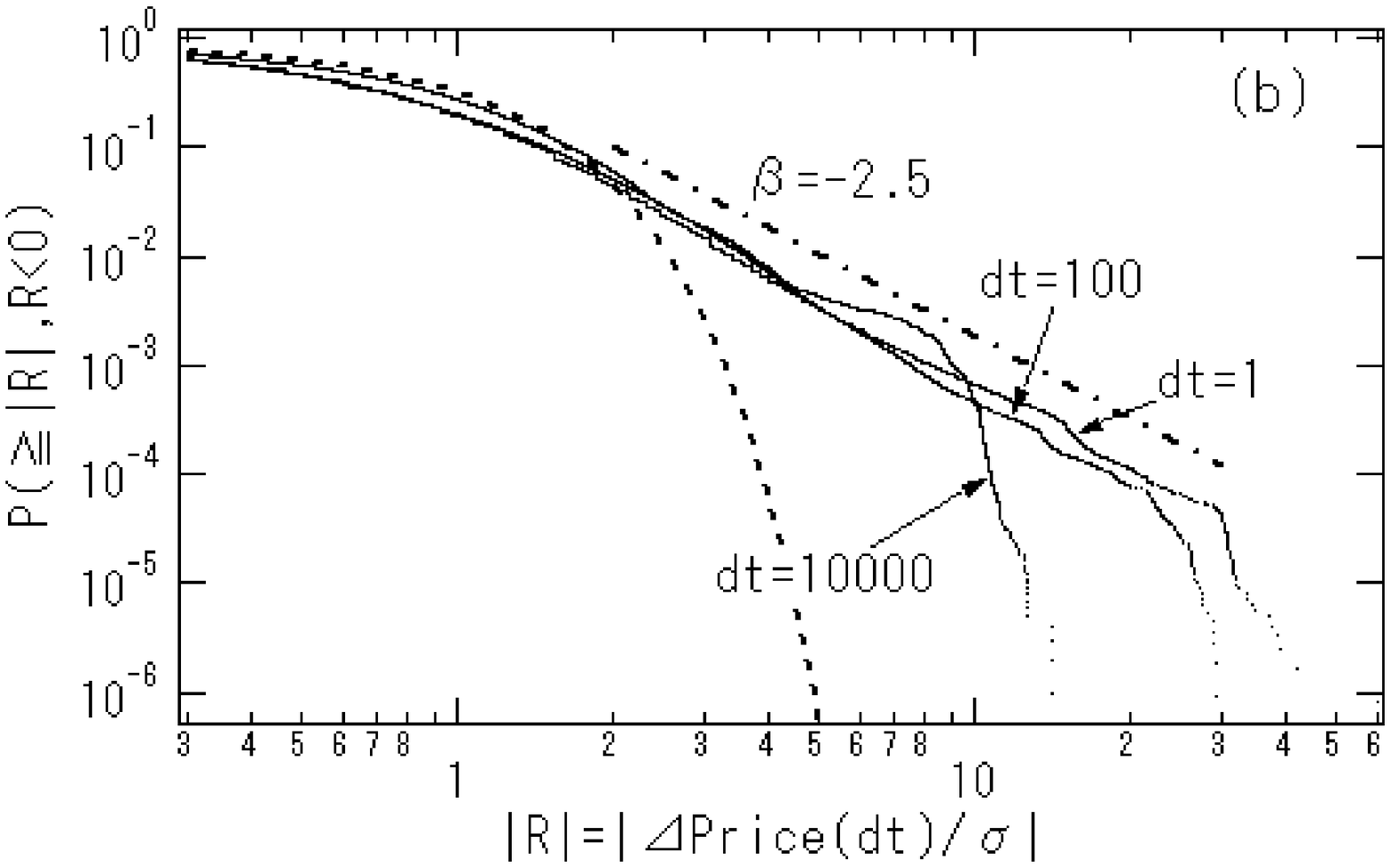}}
\label{fig2}
\end{minipage}
\textbf{Fig.1a,b} The distributions of fluctuation with 
(a) the weaker-yen, 
(b) the stronger-yen. Cumulative probability distributions of fluctuation of Yen/Dollar rate. The time scale (\textit{dt}) is 1 minute scale. 
$dt=10000$ is about 1 week. The dashed lines 
represent the straight line of the power-law with exponent $-2.5$ and 
standard normal distribution.
\end{figure}
\end{center}

\begin{center}
\begin{figure}
\begin{minipage}{.5\linewidth}
\centerline{\includegraphics[width=2.6in,height=1.56in]{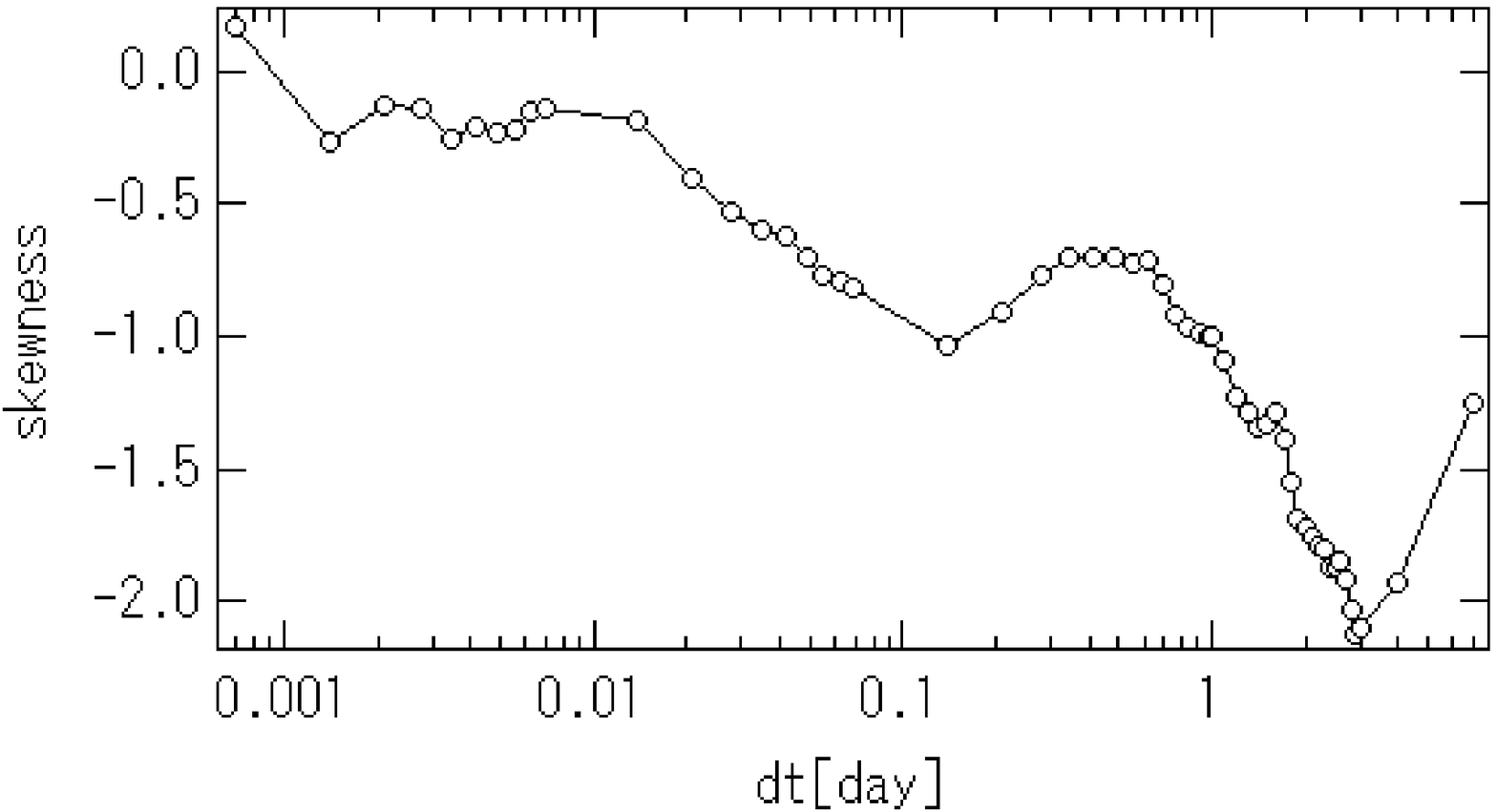}}
\label{fig3}
\textbf{Fig.2 }Skewness of the rate change \\ fluctuation.
\end{minipage}
\begin{minipage}{.5\linewidth}
\centerline{\includegraphics[width=2.55in,height=1.33in]{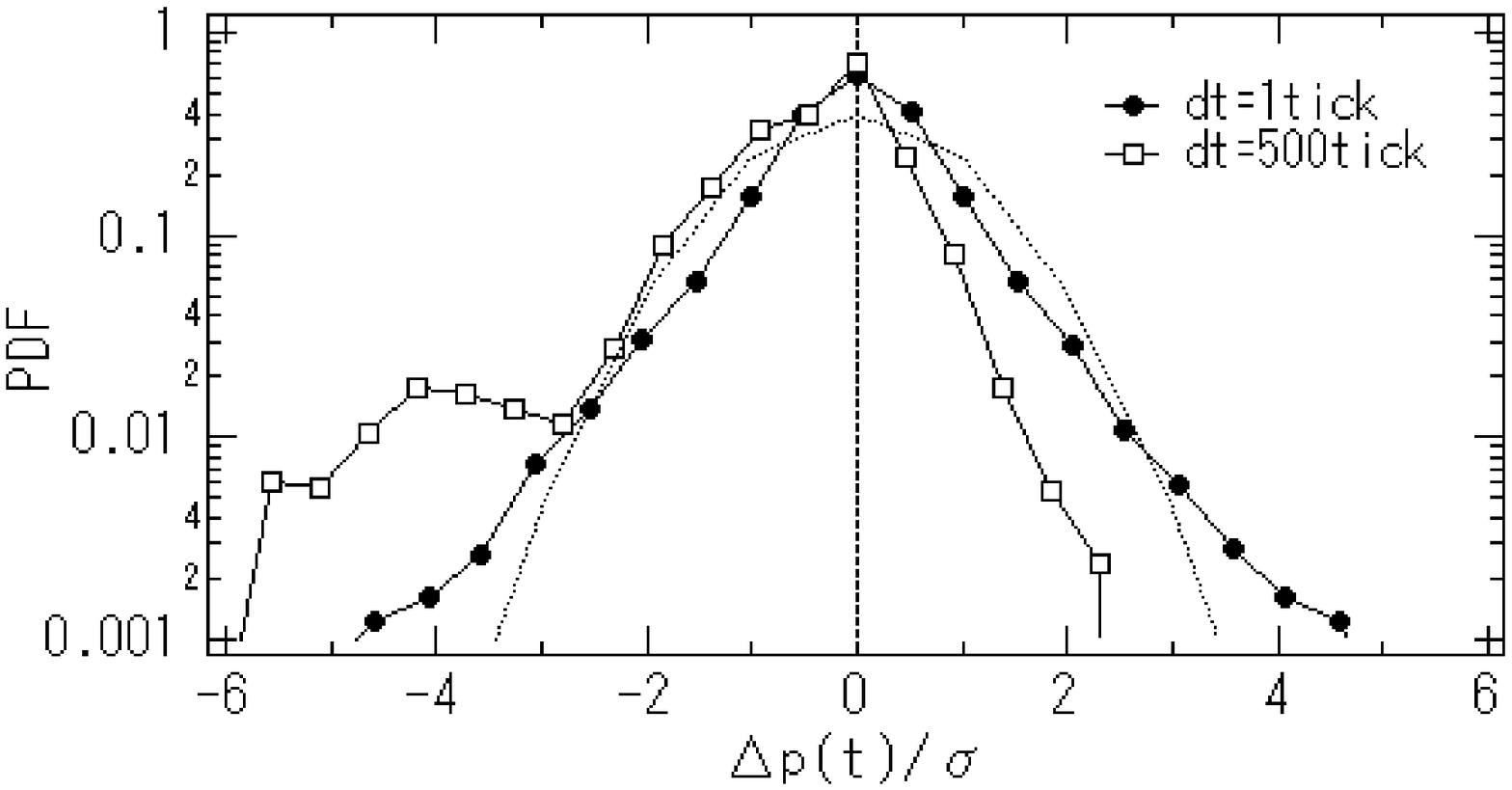}}
\label{fig4}
\textbf{Fig.3 }Probability density of rate change observed in the stronger-yen
 period of 5/10/1998 - 8/10/1998. The dashed curve 
shows the normal distribution.
\end{minipage}
\end{figure}
\end{center}

\begin{figure}[htbp]
\centerline{\includegraphics[width=3.2in,height=1.64in]{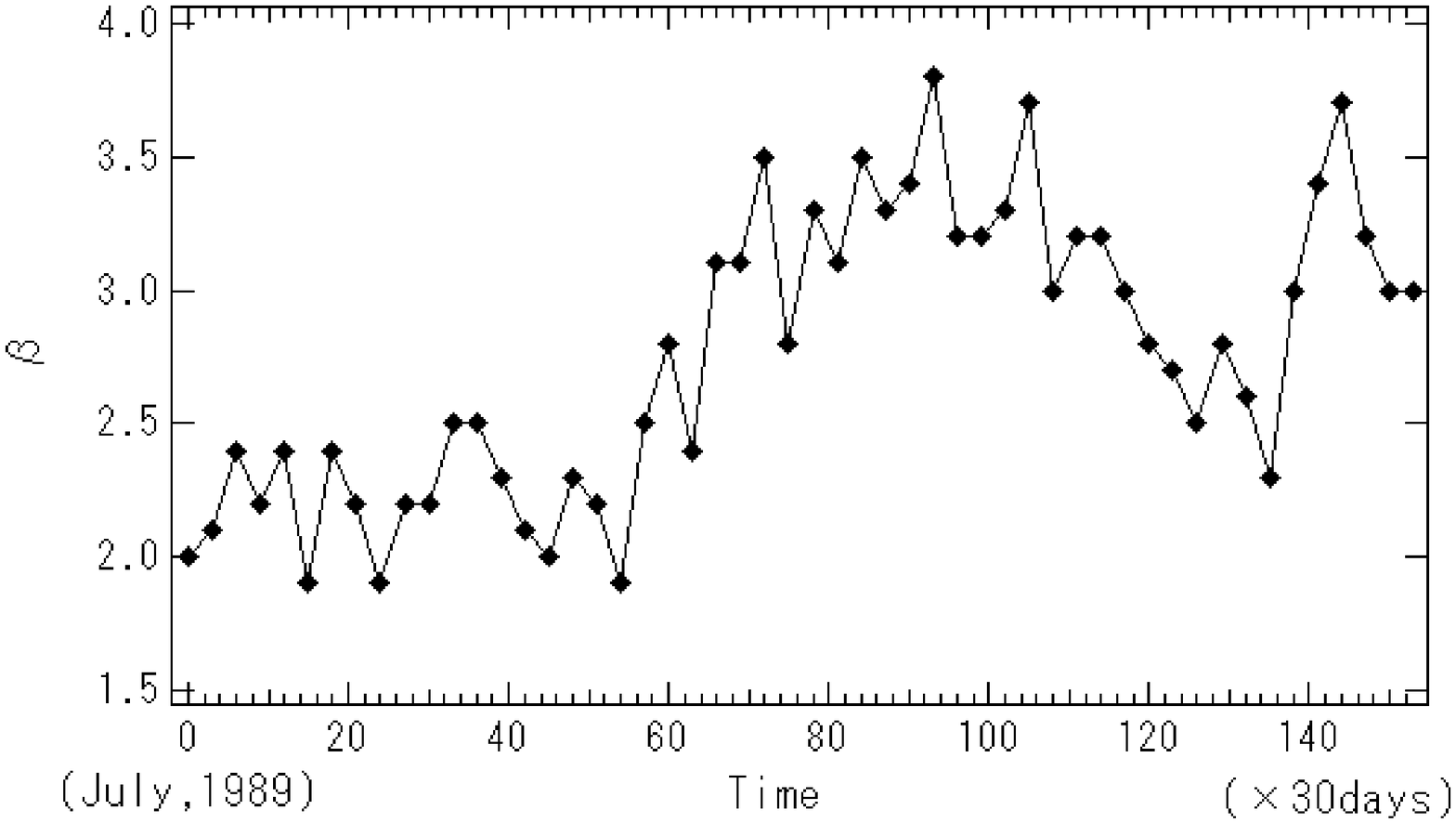}}
\label{fig5}
\textbf{Fig.4 }Time series of the distribution exponent.
The transition of distribution exponent ($\beta )$. The starting point ($t=
1$) is the exponent for July 1989.
\end{figure}

\vspace{-2cm}
In October, 1998, the dollar rate changed from 140 yen to 119 yen, about 20 
{\%} in 4 days and violent fluctuations followed for a while after this 
event. This large rate change is special. In Fig.3, the rate change 
distribution in the 4 days is shown. The distribution observed with $dt=500$ 
ticks (about 1 hour) is quite asymmetric. However, when observed with 
$dt=1$ tick, the distribution is almost symmetric. From this extreme example, 
we believe short time statistics of exchange rate is always symmetric, and 
the asymmetric properties appear for larger time scales.

We discuss about the power exponent of the price change distribution of the 
short time scale. The power exponent of distribution shows the market 
characteristics, but the power exponent of each market is different [4]. As 
shown in Fig.1, the power exponent of Yen/Dollar rate for 13 years is about 
$-2.5$. Because data covers a long term, the characteristic of 
the market is not 
uniform we think. We investigate time change of a market by estimating a 
locally defined power exponent $\beta $ in each month observing the date 
with $dt= 1$ minute. In Fig.4, we show the time series of the distribution 
exponent $\beta $. As known from this figure we find the exponent $\beta $ 
is slowly changing in the range from $-1.9$ to $-3.8$. Hence, it is impossible 
to assume that the value of the exponent is a universal constant; the 
market's conditions have been changing slowly.

\section{Correlation}
We focus on the predictability of the rate changes. The autocorrelation of 
the rate changes is defined by the following equation,

\begin{equation}
\label{eq2}
C(dt) = \frac{\left\langle {dP(t) \cdot dP(t + dt)} \right\rangle - 
\left\langle {dP(t)} \right\rangle \left\langle {dP(t + dt)} \right\rangle 
}{\sigma ^2},
\end{equation}

where, $dP(t) = P(t + 1\mbox{tick}) - P(t)$. This correlation vanishes after 
$dt=2$ ticks (about 10 seconds) [5]. Therefore, it is difficult to 
predict the future by this correlation function. However, higher order 
correlations are expected to exist in the rate changes because dealers are 
estimating the future by some ways. Although the volatility ($=\vert 
P(t+dt)-P(t)\vert$) cannot predict whether price goes up or down, long 
correlation is observed [4]. We investigate how long this correlation 
continues. For the volatility we use the absolute value of the rate change 
for 1 minute and calculate the correlation function of the volatility. As 
shown in Fig.5, we find that the correlation of volatility continues for 
about 3 months following a power law with exponent about $-0.2$. This result 
may imply that the market conditions are changing slowly with time scale of 
3 months, a season.

\begin{center}
\begin{figure}
\begin{minipage}{.5\linewidth}
\centerline{\includegraphics[width=2.52in,height=1.54in]{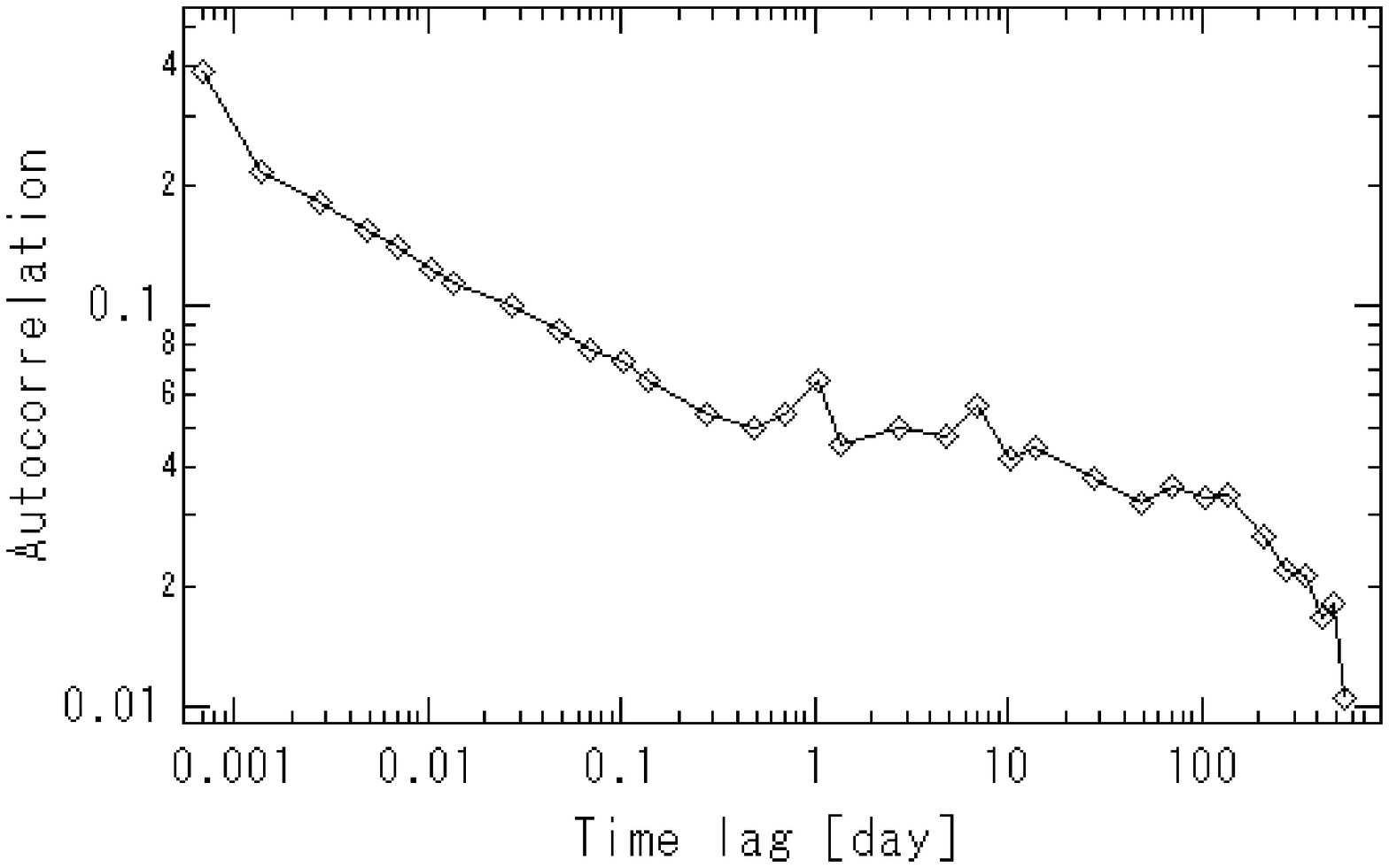}}
\label{fig6}
\textbf{Fig.5 } Autocorrelation of volatility.
\end{minipage}
\begin{minipage}{.5\linewidth}
\centerline{\includegraphics[width=2.52in,height=1.5in]{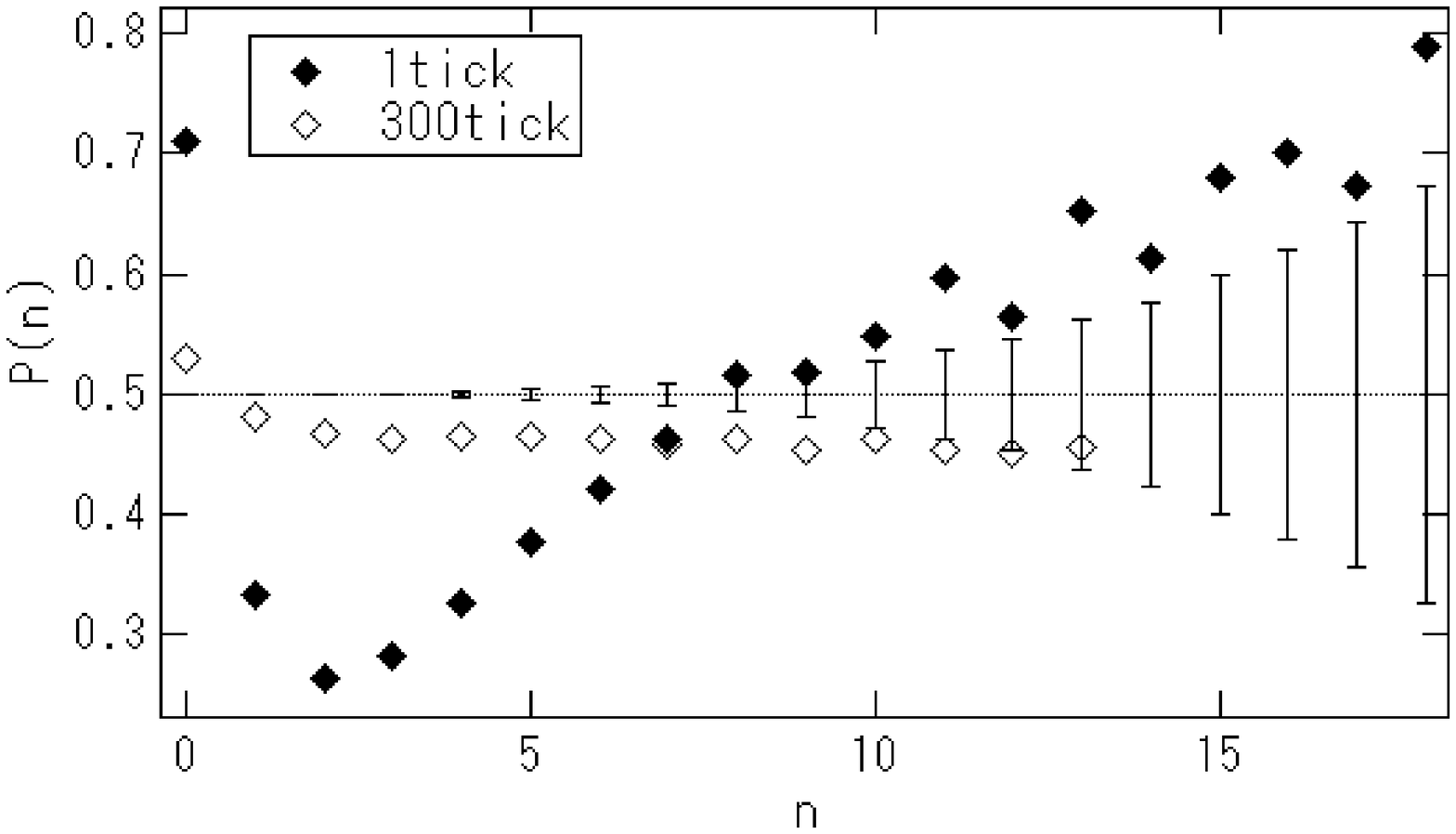}}
\label{fig7}
\textbf{Fig.6 }The conditional probability of finding $+$ after 
n-successive $+$. 
Bars show statistical errors.
\end{minipage}
\end{figure}
\end{center}

Next, we investigate higher order correlation by considering the dealer's 
strategies. We assume there are basically two strategies in dealers. The 
first strategy is so-called against-the-trend; the Fundamentalists' believe 
in the equilibrium exchange rate where market rates should move around, so 
that they trend to respond against local trends. The other strategy is the 
trend followers' who try to ride on the trends. We examine which strategy 
dominates the market from the data. To this aim we introduce the up-down 
analysis [6]. The up-down signal of "$+$" is shown when the exchange rate goes 
up and the up-down signal of "$-$" is shown when the exchange rate goes down. 
The even cases are neglected. We observe the existence of "trends" with the 
help of the conditional probability after n-successive signs. For example, 
the conditional probability after 3-successive signs is P($+\vert +++-$) or 
P($-\vert ---+$). The conditional probability distributions of $+$ and $-$ are 
almost symmetric. In Fig.6 we show conditional probability after 
n-successive signs. When the time unit is about 30 minutes, the conditional 
probability is always about 0.5, namely, the fluctuations can be regarded 
roughly as random. Now in the case of 1 tick, if the number of succession n 
is 7 ticks or less, the joint probability shows the moves against trends are 
more frequent than random cases. This shows that the strategy of 
against-the-trend is stronger in very short time scale. When n is 8 ticks or 
more, we can find the tendency that the same sign will appear with 
probability larger than 0.5, namely, the trend-followers become dominant. 
From this analysis we find that the existence of trend can be judged by 8 
ticks (or about a minute), namely, when the exchange rates move 
monotonically for more than a minute, the dealers may feel the existence of 
the trend and a large rate change will be resulted.

\section{Discussion}
Many statistical laws of the foreign exchange rates have been confirmed in 
the analysis with high precision. In the next step, we need to establish the 
dynamics which satisfies these statistical laws of microscopic time scales 
from 1 tick up to several hundred minutes scales. A set of market price 
equations has already been proposed [7] and by extending this set of 
equations we have derived the basic price equation that is valid also for 
much longer time scales including the inflations [8]. Deriving a macroscopic 
price dynamics from the statistics of microscopic fluctuations should be a 
challenging topic in the near future. We hope that the statistical laws in 
the microscopic time scales will be helpful to the stabilization of the 
world economy.

\textbf{Acknowledgments}

The present authors would like to appreciate Hiroyuki Moriya in Oxford 
Financial Education for providing us with the CQG data of high-frequency 
exchange rate, Prof. Tohru Nakano for stimulus discussions.

% \appendix

% \section{}
% \label{}

% Bibliographic references with the natbib package:
% Parenthetical: \citep{Bai92} produces (Bailyn 1992).
% Textual: \citet{Bai95} produces Bailyn et al. (1995).
% An affix and part of a reference:
%   \citep[e.g.][Ch. 2]{Bar76}
%   produces (e.g. Barnes et al. 1976, Ch. 2).

\end{document}